\documentclass[aps,prl,twocolumn,groupedaddress,superscriptaddress]{revtex4-1}

\usepackage{graphicx}
\usepackage{textcomp}
\usepackage{gensymb}
\usepackage{amsmath}
\usepackage{hyperref}
\usepackage{braket}
\usepackage{natbib}
\usepackage[utf8]{inputenc}

\begin{document}

\title{Large flux-mediated coupling in hybrid electromechanical system with a transmon qubit}

\author{Tanmoy~Bera}

\author{Sourav~Majumder}

\author{Sudhir~Kumar~Sahu}

\author{Vibhor~Singh}
\email{v.singh@iisc.ac.in} 
\affiliation{Department of Physics, Indian Institute of Science, Bangalore-560012 (India)}

\date{\today}

\begin{abstract}

Control over the quantum states of a massive oscillator is important for 
several technological applications and to test the fundamental limits 
of quantum mechanics. Addition of an internal degree of freedom to the 
oscillator could be a valuable resource for such control. Recently, hybrid 
electromechanical systems using superconducting qubits, based on electric-charge 
mediated  coupling, have been quite successful.  Here, we realize a hybrid 
device, consisting of a  superconducting transmon  qubit and a mechanical 
resonator coupled using the magnetic-flux. The coupling stems from the 
quantum-interference of the superconducting phase across the tunnel junctions. 
We demonstrate a vacuum electromechanical coupling rate up to 4~kHz by 
making the transmon qubit resonant with the readout cavity. 
Consequently, thermal-motion of the mechanical resonator is 
detected by driving the hybridized-mode with mean-occupancy 
well below one photon. By tuning qubit away from the cavity, 
electromechanical coupling can be enhanced to 40~kHz. In this 
limit, a small coherent drive on the mechanical resonator results 
in the splitting of qubit spectrum, and we observe interference 
signature arising from the Landau–Zener–Stückelberg effect.  
With improvements in qubit coherence, this system offers a novel 
platform to realize rich interactions and could potentially 
provide full control over the quantum motional states.

\end{abstract}


\maketitle

Cavity optomechanical systems, where a mechanical mode parametrically modulates the 
resonant frequency of an electromagnetic (EM) mode, have been very successful in 
controlling the motional states of massive oscillators \cite{aspelmeyer_cavity_2014}. 
Starting from the earlier demonstration of the motional quantum ground state
by the sideband cooling technique \cite{teufel_sideband_2011,chan_laser_2011}, these 
experiments have reached several milestones related to the displacement-detection 
\cite{anetsberger_measuring_2010} and the
preparation of the non-classical states of mechanical motion \cite{pirkkalainen_squeezing_2015,wollman_quantum_2015}.  
Beyond the traditional two-mode systems, consisting of one EM and one mechanical mode, 
cavity optomechanical systems with an auxiliary mode provides a wide range of interactions.
Such systems have been used to realize nonreciprocal devices \cite{peterson_demonstration_2017,bernier_nonreciprocal_2017,barzanjeh_mechanical_2017},  
and to demonstrate quantum entanglement between two mechanical resonators \cite{riedinger_remote_2018,
ockeloen-korppi_stabilized_2018}.

Among the two-mode cavity optomechanical devices, preparation of 
the quantum states of motion appears to be technologically challenging. 
One successful strategy in the microwave domain, to circumvent this, 
is to introduce an auxiliary nonlinear mode such as a qubit. 
The qubit can be used as a single-photon source \cite{reed_faithful_2017}, 
photon-counter \cite{lecocq_resolving_2015}, 
or directly coupled to a mechanical mode using its charge dispersion \cite{pirkkalainen_hybrid_2013,pirkkalainen_cavity_2015,viennot_phonon-number-sensitive_2018,sletten_resolving_2019} or 
the piezo-electric effect \cite{oconnell_quantum_2010,chu_creation_2018,arrangoiz-arriola_resolving_2019}.
In such devices, the qubit mode ‘acts’ like an additional degree 
of freedom which couples to mechanical mode via an intermediate 
mode or directly using the ``charge" dispersion.

While systems utilizing the charge-based coupling have been 
studied extensively, the experimental progress of the hybrid 
systems based on magnetic-flux has been very limited. 
Here we design and study the performance of a hybrid 
electromechanical device based on a fundamentally different 
coupling scheme based on the magnetic flux. We engineer the 
device parameters such that in addition to the flux-based 
electromechanical coupling, one mode maintains sufficient 
anharmonicity to be qualified as a qubit. This approach 
results in an electromechanical system with an internal 
spin-half degree of freedom. While, the strong and tunable 
nonlinearity of the qubit mode improves the displacement sensitivity, 
the large electromechanical coupling also manifests in the modification 
of qubit spectrum in the dispersive limit.
Similar to vacuum-electromechanical coupling rate's scaling with 
total charge in charge-dispersion based schemes \cite{pirkkalainen_hybrid_2013,pirkkalainen_cavity_2015,viennot_phonon-number-sensitive_2018}, 
the coupling rate with flux-based scheme is expected 
to scale linearly with the magnetic field. 
Therefore, such an approach has the potential to reach 
the elusive single-photon strong coupling 
regime with suitable choice of materials \cite{kounalakis_flux-mediated_2020}.

\begin{figure*}
	\includegraphics[width = 160 mm]{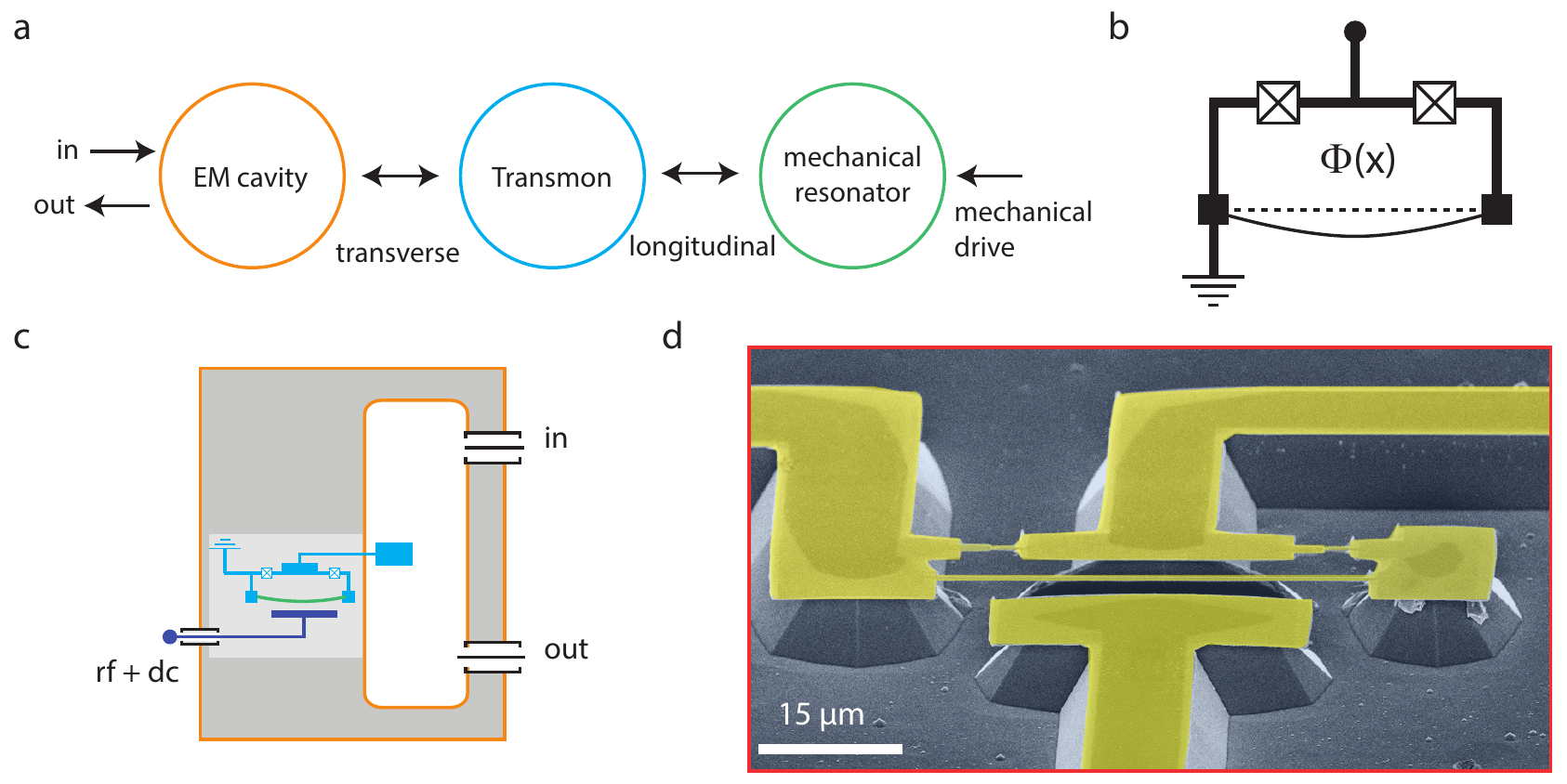}
	\caption{(a) A schematic showing various components of the hybrid electromechanical system.
		The transmon qubit couples to an electromagnetic (EM) cavity via the transverse 
		coupling. A low-frequency mechanical resonator couples to the transmon qubit via
		the longitudinal coupling. (b) A schematic of the SQUID loop with a suspended arm.
		Due to the magnetic flux $\Phi(x)$ dependence of Josephson inductance, it forms a
		displacement-dependent inductor. 
		(c) A cross-sectional view of a 3D-cavity based transmon device. The gray (white) portion represents
		the copper (machined chamber).
		Input-output ports for microwave and a third port added for mechanical actuation is 
		shown. The SQUID loop is placed inside a small recess of the cavity schematically shown by the lighter gray area.
		(d) A false-color SEM image of the SQUID loop, showing the suspended portion of the
		Josephson junctions and the nanobeam. The mechanical resonator 
		has a length and width of 45~$\mu$m and 300~nm, respectively. It consists of a 50~nm coating of 
		aluminum over 100~nm thick highly-stressed SiN film. The $T$-shaped electrode
		in the lower-half of the image is used to actuate the mechanical resonator.}
	\label{fig1}
\end{figure*}

The hybrid device consists of a transmon qubit coupled 
to a mechanical resonator and a readout cavity, as shown in Fig.~\ref{fig1}(a).
The transmon qubit couples to the cavity via a dipole coupling, commonly referred to
as transverse coupling, as it connects the ground and the first excited state of the 
qubit \cite{koch_charge-insensitive_2007}.
The mechanical resonator couples to the qubit via a flux-mediated coupling. 
Such coupling is achieved by embedding a mechanical resonator 
into one of the arms of a SQUID (Superconducting Quantum Interference Device) 
loop, which provides the necessary Josephson inductance to form a transmon qubit. 
Due to the quantum interference of the superconducting phase, 
the Josephson inductance of the SQUID depends 
on the magnetic flux threading the loop as schematically shown in Fig.~\ref{fig1}(b).
In the presence of a magnetic field applied normal to the plane of the SQUID, 
it acts like a displacement-dependent nonlinear inductor.
By shunting the SQUID ``inductor" to a suitable capacitance, 
a transmon qubit mode can be designed. 
As the motion of the mechanical resonator directly affects the qubit transition
frequency, this coupling is referred to as longitudinal coupling.
A flux-coupled hybrid system formed this way can be thought of a dual to 
the ``charge" coupling approach realized with the CPB qubit \cite{viennot_phonon-number-sensitive_2018}. 
Theoretically, the flux-mediated electromechanical coupling has been considered in the context of 
flux-qubits \cite{xue_controllable_2007}, cavity-electromechanical devices \cite{nation_ultrastrong_2016},
and more recently with the transmon qubit \cite{khosla_displacemon_2018,kounalakis_flux-mediated_2020}.
On the experimental side, the scheme has been used for large bandwidth displacement detection
\cite{etaki_motion_2008}, and to demonstrate the cavity-electromechanical system by embedding 
Josephson elements in the microwave circuitry \cite{rodrigues_coupling_2019,schmidt_sideband-resolved_2019}.

In comparison to existing flux-coupling approaches\cite{rodrigues_coupling_2019,schmidt_sideband-resolved_2019}, 
our design methodology
circumvents several issues by using a tunable transmon-mode. First, the 
requirement of large Josephson inductance for transmon design helps in suppressing hysteretic 
effects with magnetic flux arising from geometrical and kinetic 
inductance.
Second, our approach here is to implement a longitudinal coupling between transmon qubit 
and a mechanical resonator through the modulation of Josephson inductance.
This enables the interaction of the mechanical mode with the qubit in two distinct ways.
First, due to strong coupling between the qubit and cavity in the resonant limit, 
the mechanical motion directly couples to the hybridized-states.
Second, in the dispersive limit when the qubit is detuned far away from the cavity, a 
sufficiently large coupling between the qubit and the mechanical mode can be maintained. 
It thus provides a mean to use the qubit as an internal degree of freedom to 
the mechanical mode and further paves ways for measurement-based cooling and 
control protocols \cite{khosla_displacemon_2018,didier_fast_2015,leibfried_quantum_2003}.

\subsection{Device design}

We use a three-dimensional (3D) cavity to implement the transmon design as shown schematically 
in Fig.~\ref{fig1}(c).
Unlike the conventional 3D-transmon qubit, which couples differentially to the cavity mode,
we design a single-ended qubit mode by grounding one end of the SQUID loop to the 
cavity wall using a small wirebond \cite{barends_coherent_2013}. 
The other end of the SQUID loop extends towards the center of the cavity and provides 
the necessary qubit capacitance and coupling with the fundamental cavity mode. 
The rectangular cavity (35$\times$4$\times$35~mm$^3$) is machined using OFHC copper with the fundamental resonant 
mode TE$_{101}$ at $\omega_c\approx2\pi\times$~6~GHz.
A false-color SEM image of the SQUID loop is shown in Fig.~\ref{fig1}(d).
The nanobeam-shaped mechanical resonator, formed by 100~nm highly-stressed SiN film coated with 50~nm of Aluminum, and suspended part of the Josephson junctions can be clearly seen.
See supplementary information (SI) for details.
The offset in the SQUID position (away from the center of the cavity) in transmon design
allows us to bring a RF drive line for 
the electrostatic actuation of the mechanical resonator (see SI for design simulations).

The transmon qubit frequency $\omega_q$ is given by 
$\hbar\omega_q \approx \sqrt{8E_C E_J^0\left|\cos\left(\pi\Phi/ \Phi_0\right)\right|} - E_C$, where $E_J^0$
is the maximum Josephson energy, $E_C$ is the charging energy, $\Phi$ is the total flux threading the
SQUID loop, and $\Phi_0=h/2e$ is the magnetic flux quanta.
The tunability of qubit frequency with flux allows access to its
dispersive or resonant interaction with the cavity.
The interaction between the qubit and the cavity mode can be expressed as 
$\hbar J (\hat{a}\hat{\sigma}^+ + \hat{a}^{\dagger}\hat{\sigma}^-)$, where 
$\hat{a}$($\hat{\sigma}^-$) is the ladder operator for the cavity(qubit) mode and $J$ 
is the dipole coupling rate.
The electromechanical coupling arises from the modulation of 
qubit frequency caused by mechanical displacement.
As the qubit frequency can be tuned over a large range, it is convenient to 
define the vacuum electromechanical coupling 
rate between the qubit-cavity hybridized states $\left(\omega_{\pm}\right)$ and 
the mechanical resonator as,
\begin{equation}
g_{\pm}(\Phi) = \frac{\partial\omega_\pm(\Phi)}{\partial x} x_{zp} = \Phi G_\Phi^\pm \frac{x_{zp}}{w},
\label{gpm}
\end{equation}
where $G_\Phi^\pm = \partial\omega_\pm(\Phi)/\partial \Phi$ is the flux-responsivity, 
$x_{zp}$ is the quantum zero-point fluctuations of the mechanical resonator, and
$w$ is the effective width of the SQUID loop. 
Eq.~\ref{gpm} defines the coupling rate over the entire range of qubit frequencies.
The hybridized modes $\omega_{\pm}=\bar{\Delta}\pm\sqrt{(\Delta/2)^2+J^2}$ with
$\Delta=\omega_q-\omega_c$  $\bar{\Delta}=\left(\omega_q+\omega_c\right)/2$ 
approach the uncoupled qubit and cavity frequencies in the dispersive limit.
Restricting the coupled qubit-cavity system to single excitation subspace, in the 
resonant limit $\Delta\ll J$, the hybridized-modes essentially act like independent 
cavity optomechanical systems.
However, it is worth pointing out here that the interaction between the 
hybridized modes and the mechanical motion can be used to enhance the
quantum nonlinearity by designing $J$comparable to the mechanical frequency \cite{ludwig_enhanced_2012,bishop_nonlinear_2009}.

\subsection{Qubit spectroscopy and flux-responsivity}
\begin{figure*}
	\includegraphics[width = 160 mm]{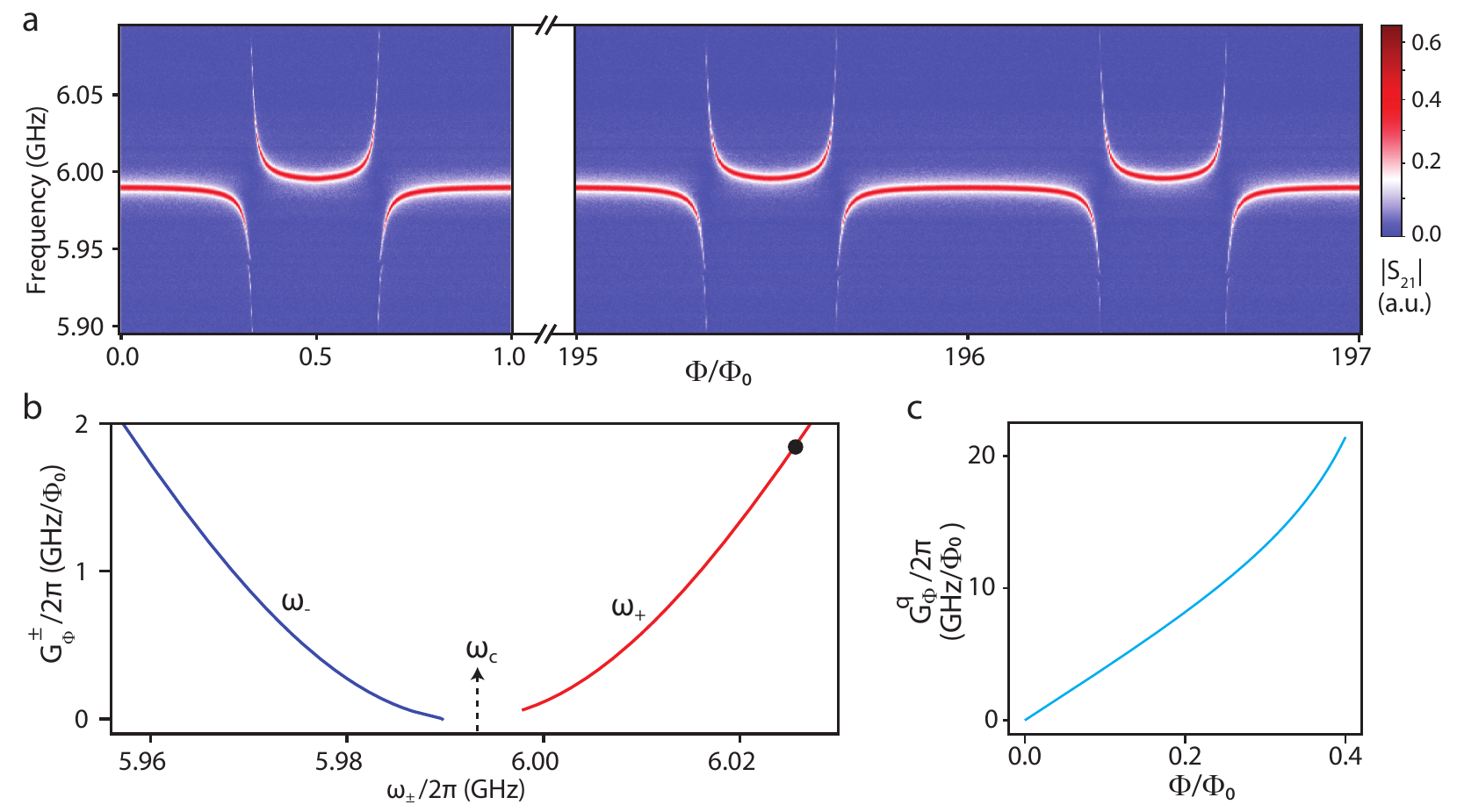}
	\caption{(a) Color scale plot of transmission $|S_{21}|$ through the cavity as the magnetic
		flux through the SQUID loop is varied. The strong qubit-cavity coupling ($J$) 
		manifests as the avoided-crossing, yielding $J\sim2\pi\times85$~MHz. 
		The extension panel shows the trend of avoided crossing at larger values of the magnetic flux.
		(b) The flux-responsivity $G^{\pm}_{\Phi} = \partial\omega_{\pm}/\partial\Phi$, computed using
		the measured device parameters, assuming identical junctions, 
		plotted as a function of hybridized frequencies. The black dot denotes the operating point of 
		6.025 GHz, corresponding to $G_{\Phi}^+/2\pi\sim$1.8~GHz/$\Phi_0$ for the subsequent measurements. 
		The arrow indicates the bare cavity frequency $\omega_c/2\pi\sim$5.993~GHz. Panel (c) 
		shows the flux-responsivity of uncoupled qubit $G^q_{\Phi} = \partial\omega_q/\partial\Phi$
		with the magnetic flux.}
	\label{fig2}
\end{figure*}

We use spectroscopic measurements to characterize 
the qubit. Fig.~\ref{fig2}(a) shows the transmission ($|S_{21}(\omega)|$) 
through the cavity as applied magnetic flux 
is varied (see SI for details of the measurement setup). When the qubit becomes resonant with the cavity, 
the vacuum Rabi splitting is observed which signifies the
strong coupling between the qubit and cavity mode. We determine a dipole 
coupling rate $J=2\pi\times$85~MHz, the bare cavity frequency $\omega_c=2\pi\times$5.993~GHz, 
maximum qubit frequency $\omega_q^0=2\pi\times$7.982~GHz, and an anharmonicity of -132~MHz (see SI for details). 
Due to the flux-periodicity of qubit frequency, the vacuum-Rabi splitting
pattern repeats with every new flux-quanta added. The extension panel of Fig.~\ref{fig2}(a)
shows the transmission measurement at a higher magnetic field. Apart from a small reduction
($\sim$15~MHz) in the maximum qubit frequency and an increase in the dressed-cavity mode linewidth, we do not
observe any significant change in the device parameters up to a field of $\sim$~3.7~mT~(310~$\Phi_0$).

To understand the flux-transduction of hybridized-modes, we compute the 
flux-responsivity $G_{\Phi}^\pm$ using the measured qubit and cavity 
parameters and assuming identical junctions.
Fig.~\ref{fig2}(b) shows the plot of $G_{\Phi}^\pm$  with 
respect to the hybridized-mode frequencies. 
The flux-responsivity of the hybridized-modes increase as their relative 
detuning $\left(|\omega_{\pm}-\omega_c|\right)$ increases.
However, reduced transmission at frequencies far away from $\omega_c$ 
hinders their use for the mechanical transduction.
We choose an optimum operating point 
of 6.025 GHz, corresponding to  $G_{\Phi}^+/2\pi\sim$1.8~GHz/$\Phi_0$, 
for the mechanical resonator characterization. This flux-responsivity is 
significantly larger than the values reported with the SQUID cavity \cite{rodrigues_coupling_2019}. 
In addition, the flux-responsivity of qubit $G^q_{\Phi}=\partial\omega_q/\partial\Phi$
can be much larger near the half-integer flux quantum as shown in 
Fig.~\ref{fig2}(c).
In the dispersive limit, while the effective coupling between the dressed 
cavity and mechanical resonator degrades by a factor of 
$\left(J/\Delta\right)^2$, a large coupling between the qubit and 
the mechanical resonator can be maintained.


\subsection{ Detection of mechanical mode and vacuum electromechanical rate}
\begin{figure*}
	\includegraphics[width = 160 mm]{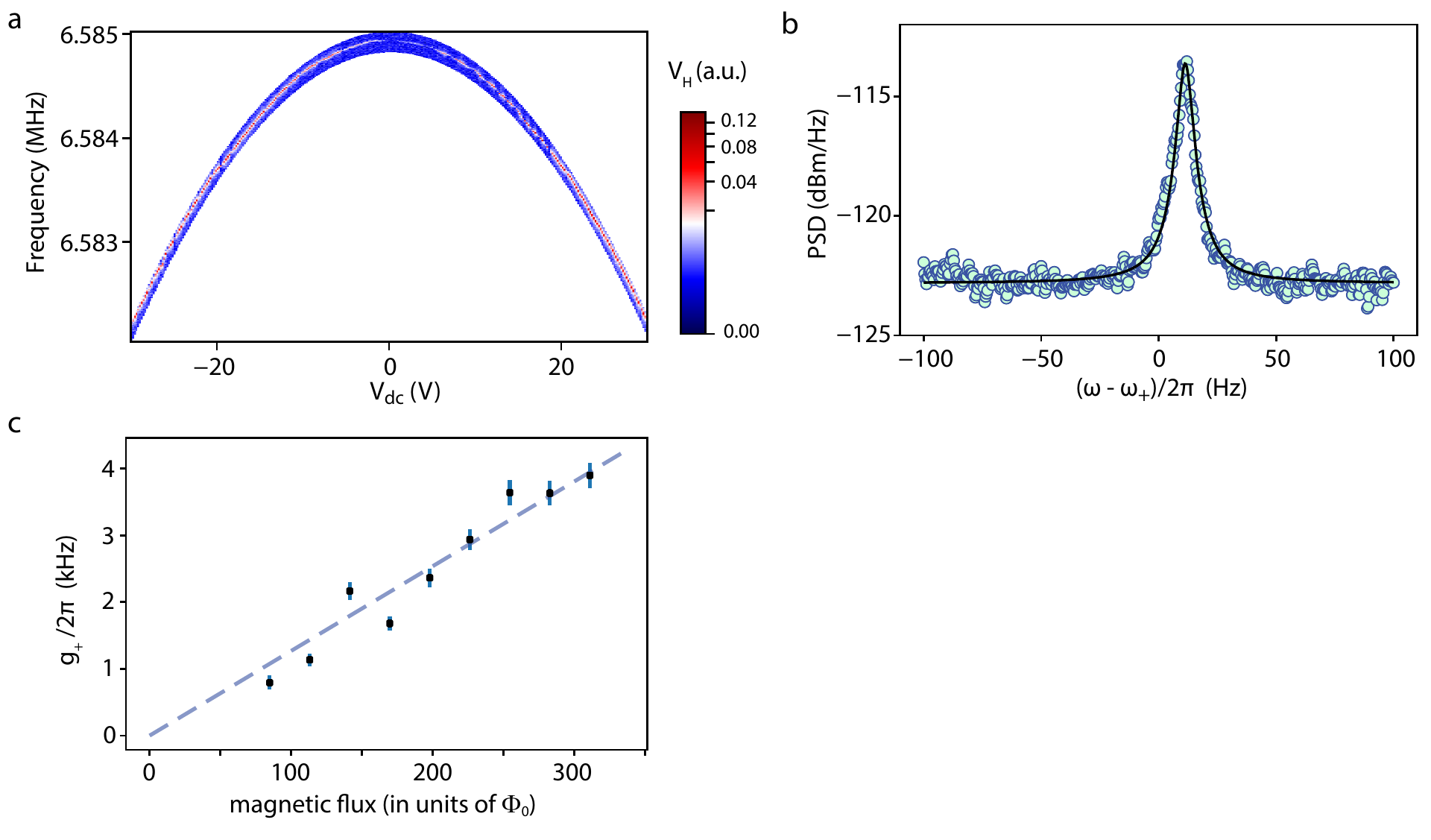}
	\caption{(a) Color plot of the mixed-down signal V$_H$ as mechanical actuation frequency and 
		dc voltage is varied. The mechanical resonance appears as a sharp change in the color.
		Blue(red) color represents low(high) values of the signal. 
		To reduce the total measurement time, the mechanical actuation frequency range 
		is automatically adjusted to follow the mechanical mode.
		(b) Average PSD along with a fitted curve yielding a mechanical linewidth $\gamma_m\sim2\pi\times6$~Hz
		corresponding to a quality factor of $\sim1.1\times10^6$.
		(c) Plot of the vacuum electromechanical coupling rate between hybridized-mode and the mechanical resonator
		as the magnetic flux through the SQUID loop is increased, while 
		$\omega_+~=~2\pi\times6.025$~GHz is kept fixed. The maximum flux applied 
		corresponds to a field of 3.7~mT. The blue-dotted line shows the expected 
		coupling rate calculated from the device parameters. The error bars represent the uncertainty resulting
		from the numerical fit of the power spectral density.}
	\label{fig3}
\end{figure*}

We first focus on the driven response of the mechanical resonator.
For electrostatic actuation, a weak ac signal and a dc voltage $V_{dc}$ 
are applied at the mechanical drive port (see SI for details). 
We fine-tune the magnetic flux near 190~$\Phi_0$ to operate the hybridized mode $\omega_+/2\pi$ at 6.025~GHz.
We inject a microwave tone at $\omega_+$ creating a mean photon 
occupation of $\approx1$, calibrated independently using ac-Stark shift.
The signal that emerges from the cavity is then mixed-down and recorded by a network analyzer. 
Fig.~\ref{fig3}(a) shows the amplitude of the signal in a color plot as the
mechanical drive frequency and $V_{dc}$ are varied.
The change in color over the background signifies the mechanical resonance.
We measure the in-plane vibrational mode at $\omega_m\sim2\pi\times$6.585~MHz
with a characteristic capacitive frequency softening with $V_{dc}$.
Next, we focus on the thermal motion of the mechanical resonator. 
We operate the hybridized mode at $\omega_+=2\pi\times$~6.025~GHz and drive the system with a microwave tone tuned 
to lower sideband ($\omega_+-\omega_m$), creating a mean photon occupation 
of $\sim$~0.1~photons.
The power spectral density (PSD) of the output signal is then recorded with a spectrum analyzer. 
The average PSD, along with the fitted Lorentzian is shown in 
Fig.~\ref{fig3}(b). We measure a mechanical linewidth of $\gamma_m=2\pi\times$6~Hz, 
corresponding to a quality factor of $\sim1.1\times$10$^6$. 
%


For a drive at the lower sideband,
the ratio of integrated power at the up-converted 
frequency $(P_m)$ near $\omega_+$ to the power of transmitted carrier signal $(P_d)$ 
at $\omega_+-\omega_m$ can be conveniently expressed as $P_m/P_d=\left(2g_{+}/\kappa\right)^2 {n}^{th}_m$,
where $n^{th}_m$ is the mean thermal occupation of the mechanical mode, and $\kappa$ is the hybridized-mode linewidth. 
By varying the fridge temperature, we estimated the mechanical mode to be thermalized to at least 50~mK or higher.
For the calculation of $g_+$, we use a thermal phonon occupancy of 169 corresponding to 53~mK (additional detail are given in SI). 
Fig.~\ref{fig3}(c) shows the variation in $g_+$ as the magnetic flux through the SQUID loop is varied,
while the hybridized-mode frequency is maintained fixed at $\omega_+/2\pi$~=~6.025~GHz.
The dotted line shows the expected electromechanical coupling rate estimated from Eq.~\ref{gpm} using 
the measured device parameters.

We emphasize that the vacuum electromechanical coupling rate 
of $g_+\sim2\pi\times$4~kHz is limited by 
choice of $\omega_+$, and the magnetic field 
range available in our measurement setup. By operating
at $\omega_c\pm J$, one can achieve the $G^+_{\Phi}/G^q_{\Phi} = 1/2$, 
resulting in $g_+\sim2\pi\times$15~kHz.
In addition, thin films of Al can withstand a larger magnetic 
field than the maximum field used here (3.7~mT). 
As the in-plane critical magnetic field is much larger than 
the perpendicular critical magnetic field for thin Al films, a configuration 
with field applied in-plane to the SQUID loop would result in 
significantly higher coupling rates for the out-of-plane mechanical mode.


\subsection{Landau-Zener-Stückelberg interference in the dispersive limit}
Next, we investigate the system by tuning 
the qubit away from $\omega_c$. In the dispersive 
limit $|\Delta|\gg J$, the mechanics essentially decouples from the cavity mode.
While the qubit-cavity interaction is given by $\left(J^2/\Delta\right)\hat{a}^\dagger\hat{a}\hat{\sigma}_z$, 
the longitudinal interaction between the uncoupled qubit and the mechanical resonator is given by 
$g_{qm}\hat{\sigma}_z(\hat{b}+\hat{b}^\dagger)$, where 
$g_{qm} = \left(\partial\omega_q/\partial x\right)x_{zp}$ is the qubit-electromechanical 
coupling rate and $\hat{b}$~$(\hat{b}^{\dagger})$ is the lowering (raising) operator for the mechanical mode. 
With a superconducting qubit device, time-dependent longitudinal coupling 
scheme has been used to perform high-fidelity qubit measurements \cite{touzard_gated_2019}.
In the present device, a static $g_{qm}$ would instead result in a small qubit-state dependent 
displacement $\left(\sim g_{qm}x_{zp}/\omega_m\right)$ \cite{didier_fast_2015}.
Here, we focus on the qubit dynamics while driving the mechanical 
resonator. The qubit is detuned to 4.9 GHz to enhance $g_{qm}$ to 40~kHz, 
and its spectrum is probed using the two-tone spectroscopy technique. 
The mechanical resonator is coherently actuated at its resonant frequency. 
It is equivalent to the flux-modulation of the qubit frequency at $\omega_m$, and a frequency deviation set by $g_{qm}$ and 
the mechanical amplitude.

\begin{figure*}
	\includegraphics[width = 95 mm]{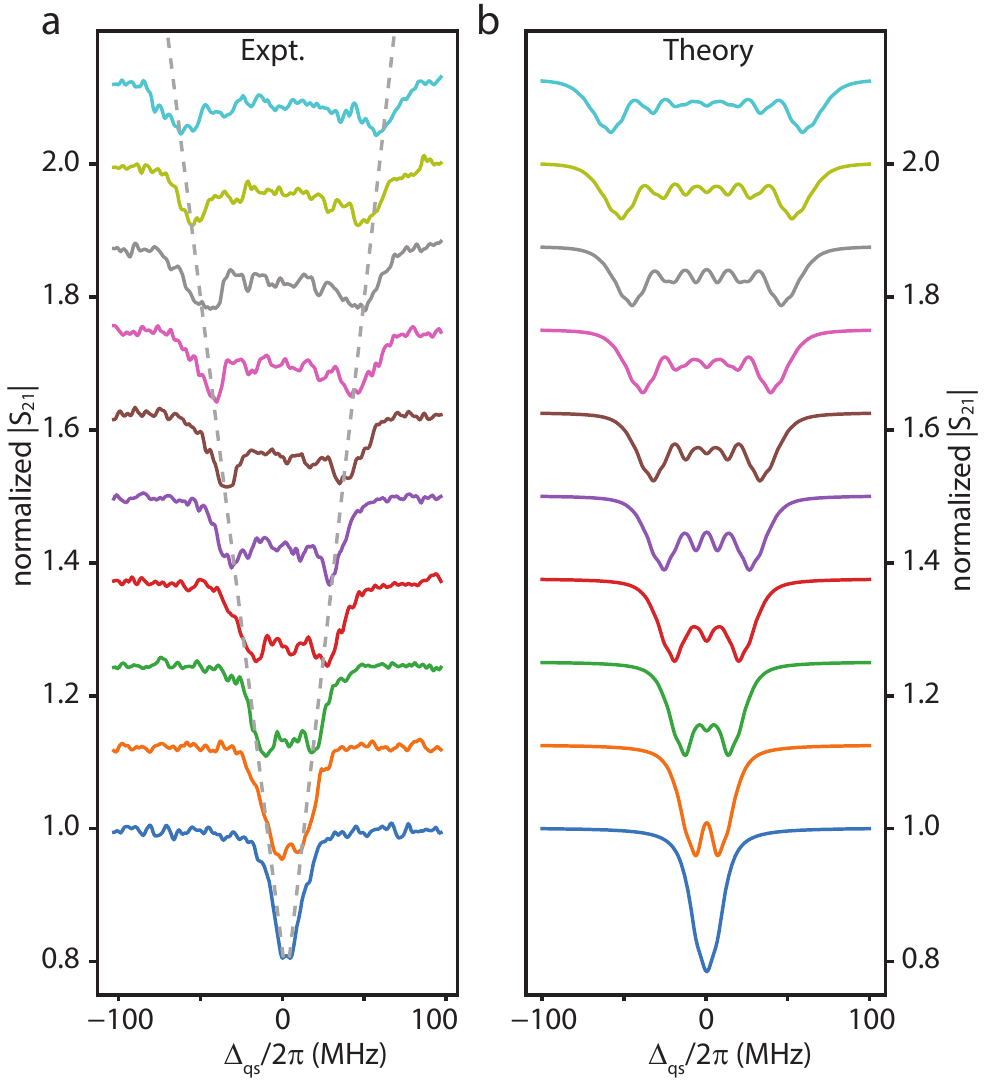}
	\caption{(a) Measurement of the qubit spectrum with the detuning of spectroscopy tone $\Delta_{qs}= \omega_s - \omega_q$, where $\omega_s$ is the spectroscopy frequency.
		The mechanical drive signal is varied from 1~to~10~mV in steps of 1~mV (bottom to top). The Rabi flop rate $\Omega_R/2\pi$ is about 3~MHz. Apart from the primary splitting, weak fringes arising from the Landau-Zener-Stückelberg interference are visible. Dotted lines are added as guide to the eye.
		(b) Qubit spectrum calculated using the master equation. The qubit frequency deviation is 
		changed from 7~MHz to 70~MHz in steps of 7~MHz (bottom to top).
		A qubit relaxation rate of 2~MHz, and a pure dephasing rate of 4~MHz is used to calculate the qubit spectrum. 
		In both panels, the probe transmission corresponding to the lowest mechanical drive is normalized to one, and 
		an offset of 0.125 has been added successively. 
	}
	\label{fig4}
\end{figure*}

Fig.~\ref{fig4}(a) shows the qubit spectrum as the strength of mechanical drive 
is varied. We observe a splitting in the qubit spectrum with a weak
modulation in-between.
The separation between the primary splitting varies linearly with the 
mechanical amplitude.
The primary splitting can be understood by considering the
passage of the system across the region of avoided crossing with separation
set by the strength of the spectroscopic tone (the Rabi-flop rate $\Omega_R$) \cite{shevchenko_landauzenerstuckelberg_2010}.
As the system moves across the avoided-crossing at a rate set by $\omega_m$, 
the transition during multiple passages mix the two states, eventually resulting in almost equal 
population of the two energy levels. Hence, it results in the splitting of the qubit
spectrum.

At large mechanical drive power, the system crosses avoided-crossing 
region with higher speed. In this regime, one would expect to see 
the interference fringes, arising 
from multiple Landau-Zener transitions, at a separation close 
to $\omega_m$ \cite{shevchenko_landauzenerstuckelberg_2010} (additional details are included in the SI).
In our experiment, as the qubit linewidth is comparable to the modulation 
frequency $\omega_m$, the fringes are not well resolved.
Their signatures are visible as weak modulation between the primary splitting.
We have performed numerical calculations based on the Lindblad master equation 
(details are included in SI). Fig.~\ref{fig4}(b) shows the result from such 
calculations. 
Apart from capturing the linear amplitude dependence of the primary splitting, 
the calculated results show the weak modulation in the experimental data.
%


In summary, we have developed  a hybrid electromechanical device by integrating a modified 
3D-transmon qubit and extremely low loss mechanical resonator of SiN/Al. 
The detection of thermo-mechanical motion by driving the system with less 
than one photon highlights the large underlying coupling rate. 
Accessibility to different regimes of interaction is further demonstrate 
by the observation of the LZS interference. 
Looking ahead, by accessing in-plane vibration mode through changes in 
the design geometry and in combination with higher magnetic field, the 
flux coupling rate can be increased a lot.
With further improvements in the coupling strength, 
the device in consideration can reach resolved sideband regime and strong 
coupling regime. This could enable experiments in the regime of the single-photon 
cooperativity exceeding one, and a conditional cooling of the mechanical 
resonator to the quantum ground state.

\subsection*{Methods}
For device fabrication, we use an intrinsic Si (100) substrate coated with 100~nm 
thick high-stress SiN layer grown using the LPCVD method. Using standard lithography and
shadow evaporation techniques, the transmon design is patterned in a single lithography
step. To release the mechanical resonator, a combination of dry and 
wet etching processes is used. 
First, the exposed SiN is vertically etched by the reactive ion etching using SF$_6$ and CHF$_3$ 
plasma. The aluminum film naturally acts as a mask layer and thus protects the SiN underneath 
it. In the second step of etching, a modified-TMAH based etchant is used to remove the 
exposed silicon, while providing excellent selectivity against Al and SiN (see SI for additional details). After the 
wet etch process, the samples are blow-dried gently with N$_2$, requiring no critical 
point drying. The (111)-facets of Si resulted from the wet etch process 
can also be seen in Fig.~\ref{fig1}(d). The sample placed inside a copper cavity, 
along with a small solenoid, is kept inside a cryoperm-shield to protect it from the ambient magnetic field 
fluctuations.

\subsection*{Acknowledgments}            
V.S. acknowledges the support received from Infosys Science Foundation, 
and under the ``Early Career Research Award" by SERB,  Department of 
Science and Technology (DST), Govt. of India. T.~B.~and S.~M. thank 
Tata Trust for providing the travel support. The authors acknowledge device 
fabrication facilities at CeNSE, IISc, Bangalore, and central facilities at 
the Department of Physics funded by DST. Authors thank R.~Vijayraghavan, Manas~Kulkarni,
Diptiman~Sen, and G.~S.~Agarwal for valuable discussions.

\subsection*{Author contribution}
T.~B. and S.~M. contributed equally to this work. 
V.~S. conceived the experiments. T.~B. and S.~M. fabricated the devices. 
T.~B., S.~M. and S.~K.~S performed the measurements. T.~B., S.~M., and V.~S. 
performed data analysis. All the authors contributed to write the 
manuscript and discuss the results.


\newpage

\end{document}


\title{Supplementary Information: Large flux-mediated coupling in hybrid electromechanical system with a transmon qubit}

\author{Tanmoy~Bera}

\author{Sourav~Majumder}

\author{Sudhir~Kumar~Sahu}

\author{Vibhor~Singh}
\email{v.singh@iisc.ac.in} 
\affiliation{Department of Physics, Indian Institute of Science, Bangalore-560012 (India)}

\maketitle

\subsection{Summary of device parameters}

Table~\ref{table-s1} lists some important device parameters.

\begin{table*}[h]
	\begin{tabular}{|c|c|c|}
		\hline 
		\multicolumn{3}{|c|}{\textbf{Device parameters}} \\
		\hline 
		
		& \textbf{Symbol}   	& \textbf{Value}  \\ \hline
		Cavity dimension & $d_x{\times}d_y{\times}d_z$ & 35$\times$35$\times$4~mm$^3$ \\ \hline
		Bare cavity frequency  & $\omega_c/2\pi$  & 5.993~GHz   \\ \hline
		Maximum qubit frequency  & $\omega_q^0/2\pi$  & 7.982~GHz  \\ \hline
		Qubit-cavity coupling rate & $J/2\pi$ & 85~MHz \\ \hline
		Anharmonicity  & $\alpha/2\pi$  & -132~MHz  \\ \hline
		Room temperature junction resistance & $R_n$  & 3.67~k$\Omega$\\ \hline
		Josephson inductance of SQUID & $L_J$ & 4.6~nH \\ \hline
		Junction capacitance & $C_J$ & 5~fF \\ \hline
		Maximum Josephson energy & $E_{J}^{0}/h$   & 33.4~GHz \\ \hline
		Charging energy from black-box simulation & $E_C/h$ & 255~MHz \\ \hline
		Cavity impedance (simulation) & $Z_c$ & 0.6~$\Omega$ \\ \hline
		Qubit impedance (simulation) & $Z_q$ & 275~$\Omega$\\ \hline
		SQUID loop area & $A$ & $\sim$~166~$\mu$m$^2$ \\ \hline
		Effective SQUID loop width (SQUID area / length of mechanical resonator) & $w$ & $\sim$~3.33~${\mu}$m \\ \hline
		SiN film thickness & $d_1$ & 100~nm \\ \hline
		Al coating thickness  & $d_2$ & 50~nm \\ \hline
		Mechanical resonator length & $l$ &  $\sim$~45~${\mu}$m \\ \hline
		Mechanical resonator width & $b$ & 300~nm\\ \hline
		Mechanical resonator thickness & $d=(d_1 + d_2)$ & 150~nm \\ \hline
		Tensile stress in SiN film  & $T$ & $\sim$~2~GPa \\ \hline
		Total mass of the mechanical resonator & $m$ & $\sim$~5.6 pg \\ \hline
		Mechanical resonator frequency  & $\omega_m/2\pi$ & 6.5849~MHz\\ \hline
		Maximum applied magnetic field & $B_{max}$ & 3.7~mT \\ \hline

	\end{tabular}
	\caption{Summary of parameters for the device studied in the main text}
	\label{table-s1}
\end{table*}

\subsection{Device Fabrication and measurement setup:}

\begin{figure}
	\includegraphics[width=140 mm]{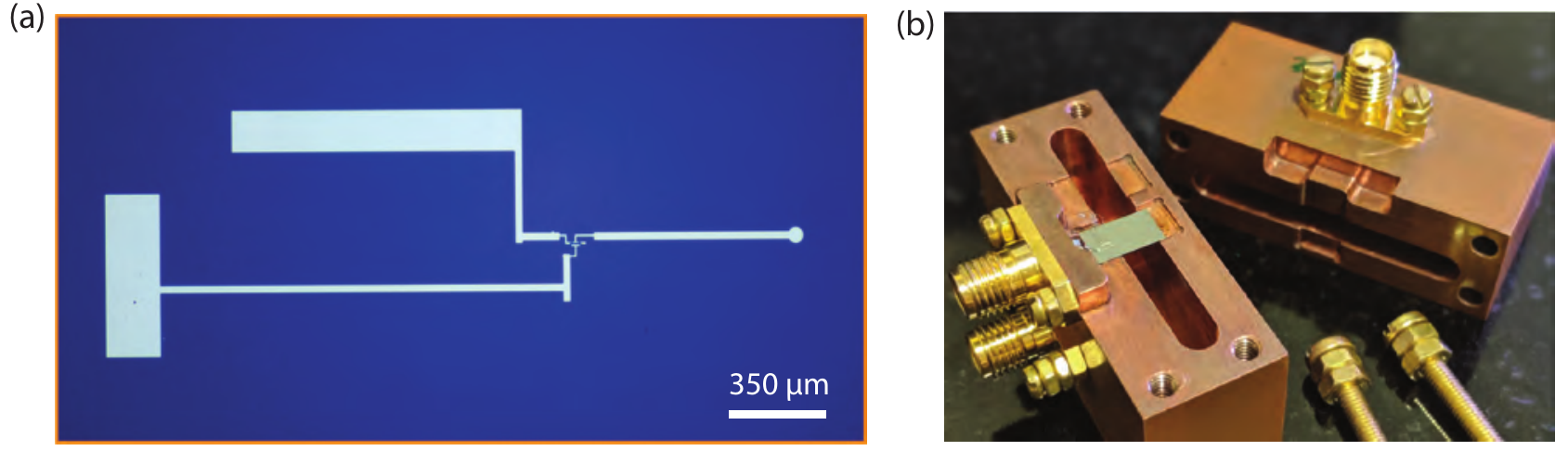}
	\caption{(a) An optical image of the device before the silicon nitride (SiN) 
		removal. The dark-blue color is due to 100~nm~SiN coating over the Si substrate. 
		(b) Image of the device placed in one half of the 3D waveguide cavity while 
		the other half is placed beside. Wirebonds used to realize a single-ended qubit 
		mode can be seen as well. The top SMA-port on the cavity-half showing two 
		connectors is used for mechanical actuation.
	}\label{fab}
\end{figure}

The device is fabricated on a 5$\times$8~mm$^2$ silicon-(100) substrate 
coated with 100~nm highly pre-stressed SiN, 
deposited using LPCVD method. The entire design is patterned in a 
single electron-beam lithography step using a bilayer resist
stack of LOR and PMMA. Subsequently, the shadow evaporation technique 
is used to deposit aluminum with an intermediate step of 
oxidation to realize tunnel Josephson junctions. Fig.~\ref{fab}(a) 
shows the optical image of the device after aluminum deposition.

To pattern the nanowire and release it from the substrate, we use two 
steps etching procedure. First, the exposed SiN is vertically etched by 
the reactive ion etching using SF$_6$ and CHF$_3$ plasma. The aluminum film 
naturally acts as a mask layer and thus protects the SiN underneath it. 
In the second step of etching, a modified TMAH based etchant is used 
to remove the exposed silicon while providing excellent selectivity against
Al and SiN \cite{yan_improved_2001,norte_nanofabrication_2015}.
For the etchant we prepare, the etch rate of Si along 100-direction is much faster than the 110- and 111-directions.
Following wet-etching, the samples are thoroughly rinsed in DI water and IPA. The 
samples are then dried using a gentle blow of N$_2$, without any critical-point drying. 
After the etch processes, we consistently observe a 30-50$\%$
increase in the room-temperature tunnel resistance of the junctions.

\begin{figure}
	\includegraphics[height=140 mm]{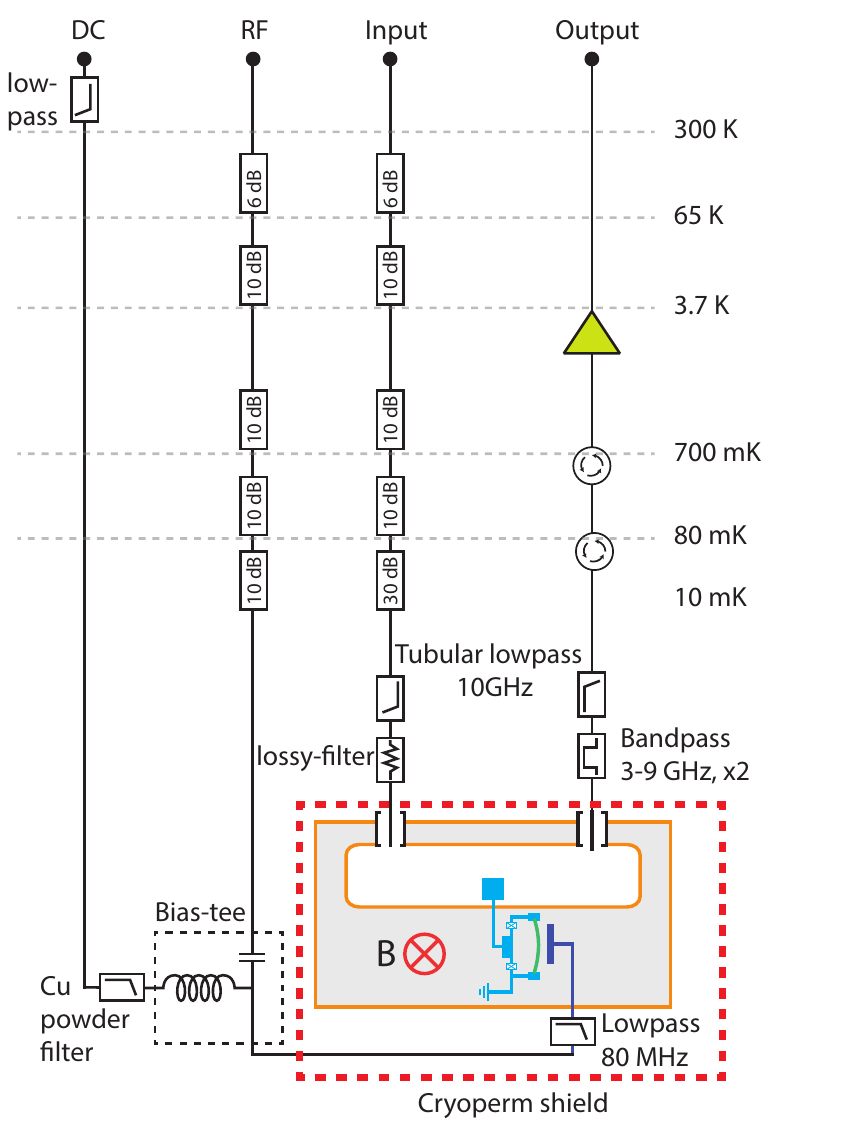}
	\caption{Schematic of the measurement setup: The input line has 
	66~dB of cryogenic attenuation, where rf line has 46 dB of fixed attenuation. 
	The bias-tee adds the rf and dc signals, which are used to actuate 
	the mechanical oscillator. The cavity is kept inside a superconducting coil, use to apply the magnetic field $B$. A 
	cryo-perm shield, shown as a red dashed rectangular box, encloses the sample and superconducting coil setup
	 and provides very effective protection against the magnetic field fluctuations outside the fridge.}
	\label{m-setup}
\end{figure}

The fabricated sample is then placed inside a 3D copper cavity, machined out of 
OFHC copper (shown in Fig.~\ref{fab}(b)). Subsequently, the cavity assembly is cooled down to 25 mK 
in a dilution refrigerator. The copper cavity, along with a small solenoid, 
is kept inside a cryoperm-shield to protect from ambient magnetic field fluctuations.
Fig.~\ref{m-setup} shows the schematic of the complete measurement setup used in the experiment.

\subsection{Device design simulation:}
\begin{figure*}
	\includegraphics[width=160 mm]{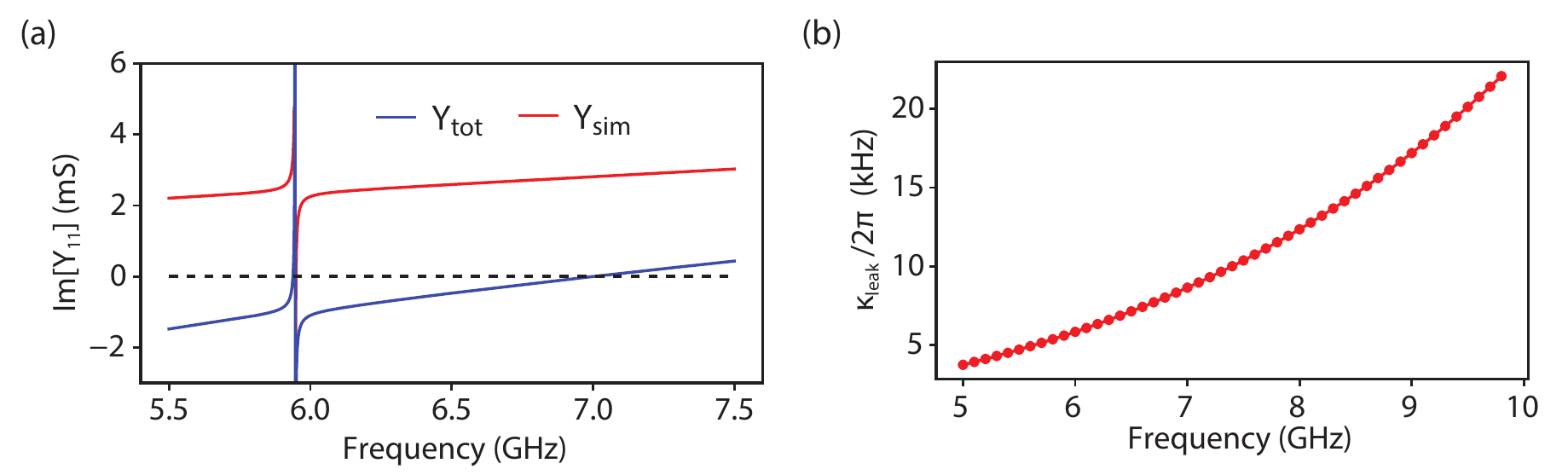}
	\caption{ (a) Admittance of the design computed using a finite element electromagnetic solver. 
		$Y_{\text{sim}}$ is the imaginary part of the admittance without the Josephson junction, and $Y_{\text{tot}}$ is 
		the imaginary part of the total admittance when the SQUID loop is present. 
		(b) The qubit relaxation rate through the mechanical actuation port 
		is plotted against the qubit frequency.}
	\label{simulation}
\end{figure*}

We use the black-box circuit quantization (BBQ) technique to simulate 
the design of the single-ended qubit\cite{nigg_black-box_2012}. We compute 
the imaginary part of the admittance $Y_{\text{sim}}$, as seen by the Josephson junction, with 
patterned substrate placed inside. For such computation, a lumped port is 
defined at the position of SQUID loop. The total reactive admittance, 
including the SQUID inductance and capacitance, is given by 
$Y_{\text{total}} = Y_{\text{sim}} + \omega C_J -\frac{1}{\omega L_J}$. Using the junction capacitance $C_J$= 5~fF 
and Josephson inductance $L_J$~=~7.5~nH of the SQUID loop, the plot of $Y_{\text{total}}$ and $Y_{\text{sim}}$ is 
shown in Fig.~\ref{simulation}(a). The zero-crossings with the positive slops in $Y_{\text{total}}$ denote the qubit and the cavity mode frequencies. By varying $L_J$, we identified the crossing on the right as the qubit mode.

The mechanical actuation electrode is designed in a way that the qubit relaxation rate through it 
can be kept lower, while maintaining sufficient actuation ability to drive the mechanical 
resonator. Apart from restricting the driveline within the cavity recess, we compute the qubit 
energy relaxation through the drive port. We simulate the qubit 
relaxation rate with and without the mechanical actuation electrode. 
The difference of two relaxation rates $k_{leak}$ as a function of qubit frequency, is shown in Fig.~\ref{simulation}(b).

\subsection{Estimation of the mechanical resonator frequency:}
The mechanical resonator is comprised of a highly pre-stressed ($\sim$2~GPa)
SiN beam of thickness 100~nm, with 50~nm coating of aluminum on top. 
Using Euler-Bernoulli's beam theory \cite{seitner_damping_2014}, 
the estimated frequency of the mechanical resonator is given by,
$$f_j = \frac{j^2 \pi}{2L^2} \sqrt{\frac{(EI)_{eff}}{(\rho A)_{eff}}} \sqrt{1+\frac{(\sigma A)_{eff}L^2}{j^2(EI)_{eff}\pi^2}}$$
where $j$~=~1, 2, 3,... denotes the vibrational mode index, $L$ is the length of 
nano-beam and $(EI)_{eff}$, $(\sigma A)_{eff}$, and $(\rho A)_{eff} $ are the 
effective tensile stress, effective bending rigidity, and effective density respectively.

The effective density is given by, 
$$(\rho A)_{eff}= \frac{\rho_{1}d_1 +\rho_{2}d_2 }{d_1 +d_2}A = \stackrel{~}{\rho }A \text{,}$$

the effective bending rigidity is given by  
$$(\sigma A)_{eff}=\frac{(\sigma_{1}d_1 +\sigma_{2}d_2)}{d_1 +d_2}A= \stackrel{~}{\sigma}A \text{,}$$
where $A = b~(d_1 + d_2)$ is the cross sectional area of the mechanical resonator. 

The effective tensile stress for in-plane and out-of-plane mode is different. 
For the out-of-plane mode, it is given by 
$$(EI)_{eff-oop}= b\frac{E_{1}^2 d_{1}^4 + 2E_1 E_2 d_2 (2d_{1}^3 + 2d_1 d_{2}^2 + 3 d_{1}^2 d_2) +E_{2}^2 d_{2}^4}{12(E_1 d_1 + E_2 d_2)}= E_1 I_{eff}^{oop} \text{.}$$ 

Similarly, for the in-plane mode, it is given by,
$$(EI)_{eff-ip} = \frac{b^3 (E_1 d_1 + E_2 d_2)}{12} = E_1 I_{eff}^{ip} \text{,}$$
where $b$ denotes the width of the mechanical resonator, $d_1$ and $d_2$ are 
the thickness of SiN and aluminium layers.

We use a Young's modulus of rigidity of $E_1$ = 160~GPa, tensile stress 
$\sigma_{1}$= 2~GPa, and mass density $\rho_{1}$ = 2800 ~kg/m$^3$ for 
SiN and $E_2$ = 69~GPa, density $\rho_{2}$ = 2700~ kg/m$^3$ for aluminum. 
The stress in aluminium film is negligible, and it does not affect the 
total tensile stress. Effectively, the aluminum coverage over SiN 
nanobeam increases the mass of the mechanical resonator, which leads to a 
decrease in the frequency. Using the parameters given above, we estimate 
the resonant frequency of the fundamental in-plane vibrational mode to be 7.7~MHz.

\subsection{Measurement of the qubit anharmonicity:}
We use the two-tone spectroscopy technique to measure the qubit anharmonicity.
A weak probe near the cavity-frequency is used to continuously monitor transmission through the cavity, while 
a second spectroscopy tone $\omega_s$ near the qubit frequency is swept. 
When $\omega_s$ matches with allowed qubit transitions,
the transmission through the cavity changes due to the dispersive coupling.

\begin{figure*}
	\includegraphics[width = 80 mm]{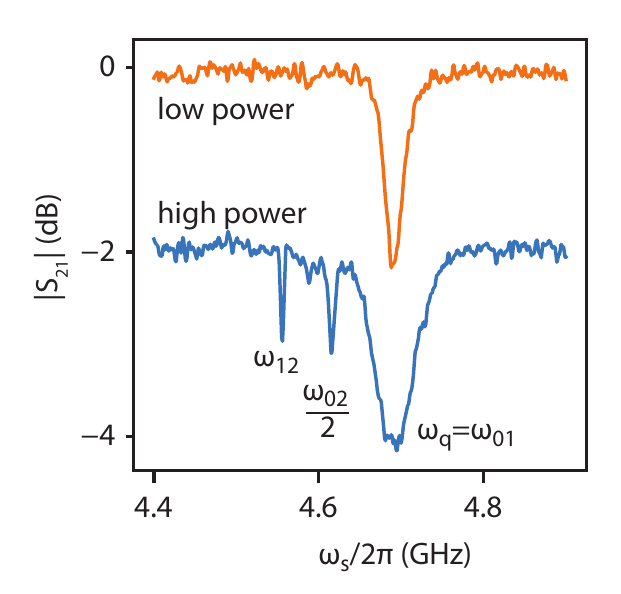}
	\caption{Two-tone spectroscopy measurement of the qubit for two different spectroscopy powers.
		At low power, the dip represents the qubit transition from ground to the first excited state.
		With strong spectroscopy drive, the higher transitions become visible. An offset of -2~dB has been added to the data at the larger drive power to bring clarity.}
	\label{anharmonic}
\end{figure*}

Fig.~\ref{anharmonic} shows the two-tone spectroscopy measurements at two different
powers of the spectroscopy tone. At larger spectroscopy power, higher 
transition $\omega_{12}$, and the two-photon transition $\omega_{02}$ become
visible. We determine the qubit anharmonicity $\alpha\equiv\omega_{12}-\omega_q$ of $-2\pi\times$132~MHz.
It is important to mention that in a traditional 3D-transmon design, the qubit
anharmonicity is approximately given by $-E_C$.
Due to the modified design used here, the geometrical inductance of the wirebonds $L_w$ dilutes the qubit
anharmonicity.

In the limit, SQUID inductance $L_J\gg L_w$, the anharmonicity is 
given by~\cite{chen_qubit_2014}, $\alpha\approx -E_C/(1 + L_w/L_{J})^3$.
From the room temperature SQUID resistance measurement and finite element simulation of $E_C$
we estimate $L_w\approx$~1.13~nH, and $L_{J}\approx$~4.6~nH.

\subsection{ac-Stark shift and the calibration of the intra-cavity photons:}

\begin{figure*}
	\includegraphics[width=130 mm]{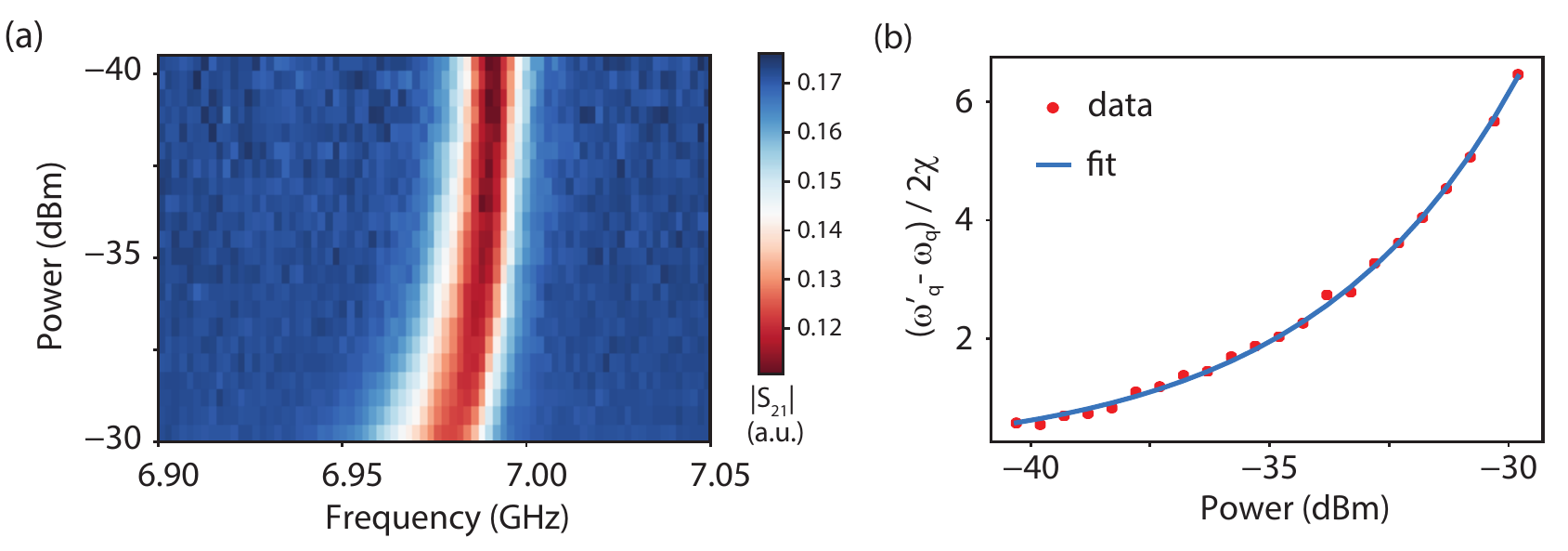}
	\caption{ (a) ac-Stark shift in two-tone spectroscopy: Plot of transmission through the cavity at dressed cavity frequency, while sweeping the spectroscopy frequency near the qubit transition with varying probe power. The red color represents the qubit transition. (b) Normalized qubit frequency shift, proportional to the number of intra-cavity photons, with increasing injected probe power.
	}\label{ac-stark}
\end{figure*}

Fig.~\ref{ac-stark}(a) shows the  two-tone spectroscopy measurement when the qubit 
is dispersively detuned. The qubit transition frequency decreases with 
increasing probe power due to the photon-induced ac-Stark effect. The shifted 
qubit frequency is given by $\omega_q'= \omega_{q} + 2n\chi$, where $n$ is the 
mean intra-cavity photon number \cite{koch_charge-insensitive_2007}.
We detune the qubit by  $\sim$~1~GHz above
the cavity mode near 6.992~GHz, and perform the two-tone spectroscopy measurements with varying probe power.

The dispersive shift is given by $\chi = J^2 
\frac{\alpha}{\Delta (\Delta + \alpha)},$ where $J$ is the coupling 
strength between the qubit and the cavity mode, $\alpha$ is the anharmonicity, and 
$\Delta = \omega_{q} - \omega_{c} $ is the qubit detuning. From the independent measurements 
of anharmonicity and qubit-cavity coupling, we compute the dispersive shift. The dispersive shift calculated this way, is then used to calibrate the number of intra-cavity photons.

The experimentally extracted 
intra-cavity photon number with increasing probe power 
from the signal generator, is plotted in Fig.~\ref{ac-stark}(b). This allows us to estimate the total microwave attenuation of the input line (from the microwave signal generator to the input port of the cavity), 
estimated to be 79.7~dB.

\subsection{Thermal motion and mechanical mode temperature calibration:}

\begin{figure*}
	\includegraphics[width=160 mm]{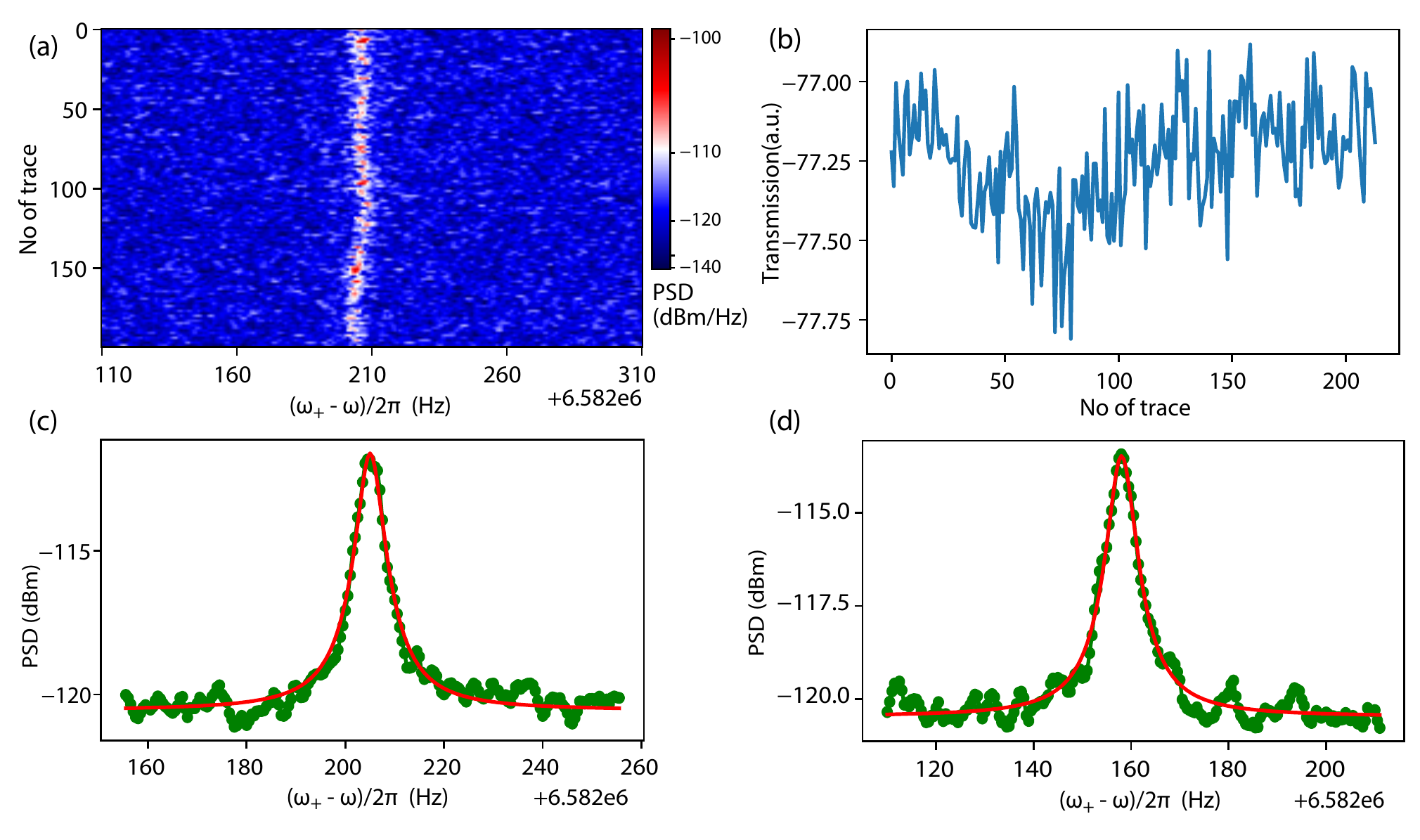}
	\caption{ (a) A color plot of 200 power spectral density traces. (b) Cavity transmission $|S_{21}(\omega_+)|$ during the measurement shown in (a). The transmission value is checked before initiating the PSD trace and validated after the PSD trace is over. 
		(c) and (d) show the average power spectral density at the lower sideband 
		measured at 25 and 50~mK, respectively. The dressed mode is driven with 
		a mean photon occupation of 0.2 at 6.025~GHz.
	}
	\label{thermal}
\end{figure*}

To determine the thermal occupation of the mechanical resonator,
we operate the system at $\omega_+/2\pi$~=~6.025~GHz. While pumping this mode with a cw-tone at 6.025~GHz, we measure the total integrated power $P_m$ in lower sideband. Also, the transmitted power~$P_d$ at the carrier frequency $\omega_+$ is recorded. 
To eliminate the records made during any flux-jump event, we record the transmission $|S_{21}|$ before and after every trace measured by the
spectrum analyzer.
While we take this precaution, it is worth pointing out that all the data shown in the main manuscript is from the measurements runs, where we did not observe any flux-jump. 
Fig.~\ref{thermal}(a) shows a 2D color-map of 200 traces of the power spectral density (PSD) measured with a spectrum analyzer. 
A plot of $|S_{21}|$ (at $\omega_+$) for all the 200 traces is shown in Fig.~\ref{thermal}(b). 

Fig.~\ref{thermal}(c) and (d) show the average trace of
lower-sideband noise spectra measured at 25~mK and 50~mK, respectively. 
The down-converted power at the lower sideband frequency for zero detuning driving is given by $P_m=P_d \left(g_{+}/(\kappa^2/4 + \omega_m^2)\right) {n}_{m}^{th}$.

\begin{table*}[h]
	\begin{tabular}{|p{8cm}|p{3cm}|p{3cm}|}
		\hline
		Device parameter    & $T$~=~25~mK 	&  $T$~=~50~mK \\ \hline
		Transmitted power at carrier frequency P$_d$  & -75.9~dBm & -77.3~dBm  \\ \hline
		Cavity dissipation rate $\kappa/2\pi$  &	 4.1~MHz  &	 5.1~MHz  \\ \hline
		
	\end{tabular}
	\caption{System parameters used for the calibration of phonon occupancy}
	\label{table-s2}
\end{table*}

Assuming the vacuum coupling rate to be the same for two temperatures, from the 
results shown in Fig.~\ref{thermal}(c), (d) and the parameter values in Table~\ref{table-s2}, 
we find the ratio $n_{m}^{25}/n_{m}^{50} \approx 1$. 
This led us to conclude that the mechanical resonator is thermalized to 50 mK or a higher temperature.
The vacuum coupling rate, shown in Fig. 3(c) of the main text, has been calculated assuming a thermalization to 53 mK as these values fall roughly on the estimated values of $g_+$ from the device parameters.

\subsection{Theoretical Model to understand Landau-Zener-Stückelberg (LSZ) interference:}

\begin{figure}
	\includegraphics[width=160 mm]{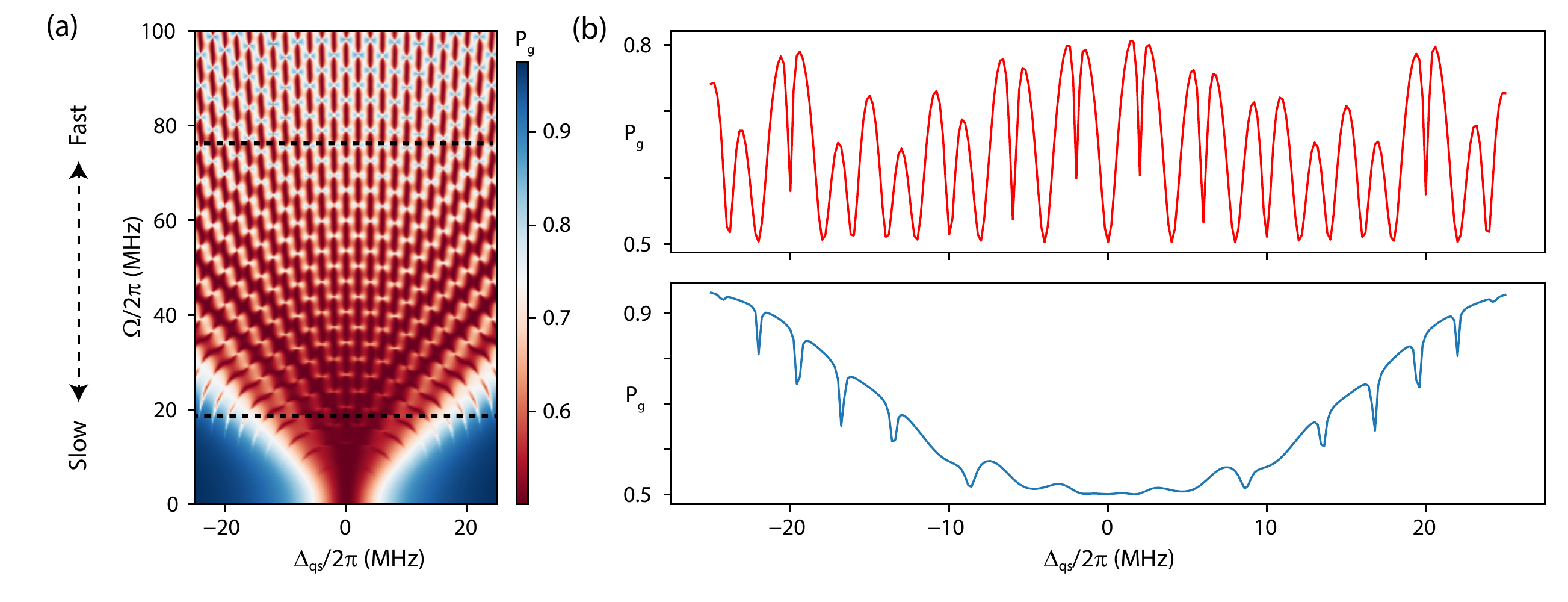}
	\caption{(a) The LZS  interference pattern with increasing qubit frequency deviation. The time-averaged value of $P_g = (1 - \langle \hat{\sigma}_{z} \rangle)/2$ is plotted with respect to detuning  $\Delta_{qs}$. (b) The top (bottom) plot shows the response for high (low) values of qubit frequency deviation. The linecuts are taken at $\Omega/2\pi = 76~$MHz (top)  and $18~$MHz (bottom), indicated by the black dashed line in the color plot. The calculations were done using $\Omega_R/2\pi = 8~$MHz, $\omega_{m}/2\pi = 2~$MHz and $\gamma_{1}/2\pi = 0.1~$MHz.}
	\label{LSZ}
\end{figure}

When the qubit is detuned far away from the cavity, the total Hamiltonian of the 
system can be written as~\cite{koch_charge-insensitive_2007},
\begin{equation}
H^{dis} = \hbar \omega_{c} \hat{a}^\dagger\hat{a} + \frac{\hbar \omega_{q}}{2}\hat{\sigma}_{z} + \hbar \omega_{m} \hat{b}^\dagger\hat{b} + \hbar\chi\hat{\sigma}_{z} \hat{a}^\dagger\hat{a} + \hbar g_{qm} \hat{\sigma}_{z}(\hat{b}+\hat{b}^\dagger)
\label{dispersive hamiltonian}
\end{equation}
Due to a large difference between the qubit and mechanical resonator 
frequency, and $g_{qm}$ being much smaller than the qubit frequency, 
we invoke the adiabatic approximation.
A coherent drive on the mechanical resonator effectively results 
in a frequency-modulated qubit. The effective Hamiltonian can be written as,

\begin{equation}
H = \hbar \omega_{c} \hat{a}^\dagger\hat{a} + \frac{\hbar (\omega_{q}+\Omega \sin(\omega_m t))}{2}\hat{\sigma}_{z} + \hbar \omega_{m} \hat{b}^\dagger\hat{b} + \hbar\chi\hat{\sigma}_{z} \hat{a}^\dagger\hat{a}
\label{driven hamiltonian}
\end{equation}
where $\Omega = g_{qm}\epsilon_{m}/x_{zp}$ is the qubit frequency 
deviation, and $\epsilon_{m}$ is the mechanical amplitude. 
Due to the longitudinal coupling, the mechanical resonator only 
contributes to the modulation of the 
qubit frequency. 

To simulate the two-tone spectroscopy data (main text Fig.4), we add probe and 
spectroscopy drives to the Hamiltonian. After performing the rotating frame 
transformation at the spectroscopy frequency $\omega_{s}$ and 
the probe frequency, we get an effective Hamiltonian given by,
\begin{equation}
H = \frac{\hbar (\Delta_{qs}+\Omega \sin(\omega_m t))}{2}\hat{\sigma}_{z} + \hbar\chi(1+\hat{\sigma}_{z}) \hat{a}^\dagger\hat{a} + \epsilon_{probe}(\hat{a} + \hat{a}^\dagger) + \frac{\Omega_R}{2}\hat{\sigma}_{x} \text{,}
\label{RWAT hamiltonian}
\end{equation}
where $\Delta_{qs} = \omega_{q} - \omega_{s}$, $\epsilon_{probe}$ is the amplitude of the probe 
signal, and $\Omega_R$ is the amplitude of the spectroscopy signal (the Rabi-flop rate).

The time evolution of different operators can be calculated by 
using the master equation solver of the QuTip package 
\cite{johansson_qutip:_2012}.
We solve for the steady-state of $\langle \hat{a} \rangle_{ss}$ using 
the total Hamiltonian and define the 
transmission as the ratio of the time-averaged value of $\langle \hat{a} \rangle_{ss}$ 
to the probe amplitude \textit{i.e.} $S_{21} = \overline{\langle \hat{a}\rangle}_{ss} / \epsilon_{probe}$.
Spectroscopy signal was varied near the qubit transition, while the probe signal frequency
was kept fixed at the dressed cavity frequency corresponding to the qubit being in 
the ground state. Their amplitudes were kept constant during simulation
\textit{i.e.} $\Omega_R/2\pi=3$~MHz and $\epsilon_{probe}/2\pi=10$~kHz.
We use a dispersive shift $\chi/2\pi=-0.71$~MHz.
The results from such calculations are plotted as the solid lines 
in Fig.~4(b) of the main text.

In the two-tone spectroscopy measurements, the measured signal
is directly related to $\langle\hat{\sigma}_{z}\rangle$. Therefore, the spectrum 
can also be worked out using the Hamiltonian of the qubit subspace only.
The model, therefore, can be simplified to a two-level system (TLS), 
which is driven along the  longitudinal direction (by the mechanical 
motion) and along the transverse direction (by the spectroscopy tone) 
simultaneously. In a frame rotating at the spectroscopy frequency, 
the Hamiltonian can be written as,
\begin{equation}
H_{TLS} = \frac{\Delta_{qs}+\Omega\sin(\omega_m t)}{2}\hat{\sigma}_{z} + \frac{\Omega_R}{2}\hat{\sigma}_{x}~\text{.}
\end{equation}

The time evolution of the system can be worked out by using the Lindblad master equation,
\begin{equation}
\dot{\rho} = -\frac{i}{\hbar}[H_{TLS},\rho] + \gamma_{1}\mathcal{D}[\hat{\sigma}_{-}]\rho + \frac{\gamma_{\phi}}{2}\mathcal{D}[\hat{\sigma}_{z}]\rho,
\label{lindbald}
\end{equation}
where $\gamma_{1}$ and $\gamma_{\phi}$ are the qubit relaxation and 
the qubit pure dephasing rates, respectively and the Lindblad superoperator~$\mathcal{D}[\hat{F}]$ is defined as,
\begin{equation}
\mathcal{D}[\hat{F}]\rho = \hat{F}\rho\hat{F}^\dagger - \frac{1}{2}\hat{F}^\dagger\hat{F}\rho - \frac{1}{2}\rho\hat{F}^\dagger\hat{F}.
\end{equation}

This leads to a set of equation of motion as:
\begin{subequations}
	\begin{equation}
	\frac{d}{dt}\langle \hat{\sigma}_{x} \rangle = -(\Delta_{qs}+\Omega \sin(\omega_m t))\langle \hat{\sigma}_{y} \rangle - (\frac{\gamma_{1}}{2}+\gamma_{\phi}) \langle \hat{\sigma}_{x} \rangle
	\end{equation}
	\begin{equation}
	\frac{d}{dt}\langle \hat{\sigma}_{y} \rangle = -\Omega_R \langle \hat{\sigma}_{z} \rangle + (\Delta_{qs}+\Omega \sin(\omega_m t)) \langle \hat{\sigma}_{x} \rangle - (\frac{\gamma_{1}}{2} + \gamma_{\phi}) \langle \hat{\sigma}_{y} \rangle
	\end{equation}
	\begin{equation}
	\frac{d}{dt}\langle \hat{\sigma}_{z} \rangle = \Omega_R \langle \hat{\sigma}_{y} \rangle - \gamma_{1} (1+\langle \hat{\sigma}_{z} \rangle)
	\end{equation}
	\label{equation of motion}
\end{subequations}

We compute the steady-state value of $\langle \hat{\sigma}_{z} \rangle$ by numerically integrating the Equation~\ref{equation of motion}. To replicate the measurements, we vary the spectroscopy frequency near the $\omega_q$ ($\Delta_{qs}$) and strength of the modulation $\Omega$ and plot the steady state value of $(1 - \langle \hat{\sigma}_{z} \rangle)/2$.

Fig.~\ref{LSZ} shows the result from the numerical calculations showing the 
characteristic Landau-Zener-Stückelberg (LZS) interference pattern \cite{shevchenko_landauzenerstuckelberg_2010}. 
In the limit of fast-passage across the avoided crossing 
$(\Omega_R\ll \Omega\omega_{m}/\Omega_R)$, the interference fringes are separated 
by the modulation frequency $\omega_{m}$.
In the slow-passage limit $(\Omega_R \gtrsim \Omega\omega_{m}/\Omega_R)$, as 
the system undergoes the avoided crossing, the probability of diabatic transition 
increases, and the separation between the fringes is no longer solely determined 
by $\omega_m$.

Splitting of the qubit spectrum can also be understand from a semi-classical model. In this approach, we calculate the 
time averaged value of 
$\langle \hat{\sigma}_{z} \rangle$ when qubit frequency is being modulated at $\omega_{m}$, thereby asserting that adiabatic approximation.

Using Eq.~\ref{equation of motion}, the steady state value of $\langle \hat{\sigma}_z \rangle$ can be obtained as

\begin{equation}
\langle \hat{\sigma}_z \rangle = - 1 + \frac{\Omega^2_R \frac{\gamma_2}{\gamma_1}}{\gamma_2^2 + \Omega^2_R \frac{\gamma_2}{\gamma_1} + [\Delta_{qs} + \Omega \sin(\omega_m t)]^2}\text{,}
\end{equation}

where $\gamma_2 = \left( \frac{\gamma_1}{2} + \gamma_{\phi} \right)$. We define the time averaged value of $\langle \hat{\sigma}_z \rangle$ as,

\begin{equation}
\overline{\langle \hat{\sigma}_z \rangle} = \frac{1}{T} \int_{0}^{T} \langle \hat{\sigma}_z \rangle~dt\text{,} 
\end{equation}

where $T = 2\pi/\omega_m$ is the time period of the mechanical oscillation. Therefore,

\begin{equation}
\overline{\langle \hat{\sigma}_z \rangle} = -1 + \frac{\Omega^2_R}{T}\frac{\gamma_2}{\gamma_1} \int_{0}^{T} \frac{dt}{\gamma_2^2 + \Omega^2_R \frac{\gamma_2}{\gamma_1} + [\Delta_{qs} + \Omega \sin(\frac{2 \pi t}{T})]^2}\text{.}
\end{equation}

To carry out the integral, we first scale the time variable and then recast the integral as,

\begin{equation}
\overline{\langle \hat{\sigma}_z \rangle} = -1  - \frac{\Omega^2_R \frac{\gamma_2}{\gamma_1}}{2 \pi \sqrt{\gamma_2^2 + \Omega^2_R \frac{\gamma_2}{\gamma_1}}} ~ \text{Im}\left[ \int_{0}^{2 \pi} \frac{dx}{i~\sqrt{\gamma_2^2 + \Omega^2_R \frac{\gamma_2}{\gamma_1}} + \Delta_{qs} + \Omega \sin(x)}\right] 
\end{equation}

This integral can be transformed into a function of a complex variable and solved using Cauchy's integral formula with a contour described by $|z| = 1$. The important part of the calculation is the fact that for any value of $\Delta_{qs}$, there is only one pole that exists inside the contour. Using the residue theorem, the final answer can be written as,

\begin{equation}
\overline{\langle \hat{\sigma}_z \rangle} = -1 + \frac{\Omega^2_R}{\beta} \frac{\gamma_2}{\gamma_1} \left|\sin\left(\frac{\theta}{2}\right)\right|
\label{sigma_z}\text{,}
\end{equation}

\begin{equation}
\beta = \sqrt{\gamma_2^2 + \Omega^2_R \frac{\gamma_2}{\gamma_1}}~ \sqrt[4]{\left(\Delta^2_{qs} - \gamma_2^2 - \Omega^2_R \frac{\gamma_2}{\gamma_1} - \Omega^2\right)^2 + 4 \Delta^2_{qs} \left(\gamma_2^2 + \Omega^2_R \frac{\gamma_2}{\gamma_1}\right)}
\end{equation}

and

\begin{equation}
\theta = \tan^{-1} \left[ \frac{2 \Delta_{qs} \sqrt{\gamma_2^2 + \Omega^2_R \frac{\gamma_2}{\gamma_1}}}{\Delta^2_{qs} - \gamma_2^2 - \Omega^2_R \frac{\gamma_2}{\gamma_1} - \Omega^2} \right]
\end{equation}

Using the decoupled cavity-Bloch equation in the steady state, the normalized transmission through the cavity can be written as,

\begin{equation}
S_{21} = \frac{-i\kappa/2}{\kappa/2 + i \chi \left(1+\overline{\langle \hat{\sigma}_z \rangle}\right)}\label{eq:norm_s21}
\end{equation}

\begin{figure}
	\includegraphics[width=50 mm]{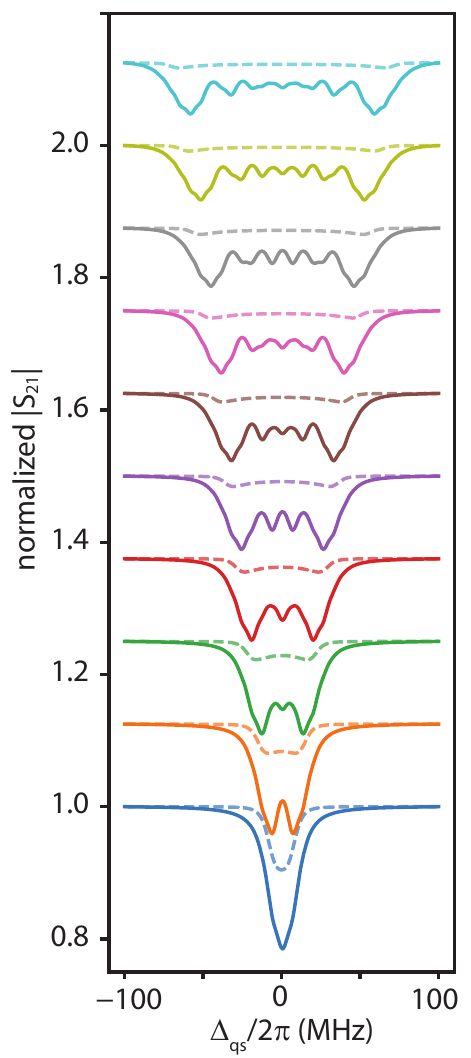}
	\caption{The solid lines show the results from the Master equation. The dotted lines show the results from semi-classical calculation. Clearly the calculations performed using the Master equation matches the experimental results better. The calculations were done using the same parameters as used for Fig-4 shown in the main text.\label{LSZ-semi}}
\end{figure}

Figure~\ref{LSZ-semi} shows the qubit spectrum using Eq.~\ref{eq:norm_s21}. For comparison, solution obtained from the Master equation have been included as well.

%